# Ferromagnetic resonance in thin ferromagnetic film with surface anisotropy

N. A. Usov[1,2]

[1]*National University of Science and Technology «MISiS», 119049, Moscow, Russia*
[2]*Pushkov Institute of Terrestrial Magnetism, Ionosphere and Radio Wave Propagation, Russian Academy of Sciences, IZMIRAN, 108480, Troitsk, Moscow, Russia*

**Abstract** The ferromagnetic resonance frequencies are obtained for a thin ferromagnetic film with surface anisotropy for the cases when the external magnetic field is applied perpendicularly or parallel to the film surface, and for various combinations of boundary conditions on the film surface. It is shown that in the presence of surface anisotropy the ferromagnetic resonance frequency essentially depends both on the film thickness and on the value of the surface anisotropy constant. The results obtained provide a basis for the correct interpretation of experimental data obtained by means of broadband ferromagnetic resonance in thin film structures.

## I. Introduction

Thin ferromagnetic films with surface magnetic anisotropy are currently of great technological interest [1-4]. In particular, the ultra-thin CoFeB/MgO bilayers [2-4] exhibit a perpendicular magnetic anisotropy. The latter provides a high thermal stability and is highly desirable for reducing the critical current for spin-torque switching [2]. Therefore, thin CoFeB ferromagnetic films are considered currently as promising electrode material in magnetic random access memory and spin-transfer torque memory devices.

From physical considerations, it is obvious that only the surface magnetic anisotropy [5-8] is capable of providing an out-of-plane magnetization in a thin ferromagnetic film made of a soft magnetic material. Actually, if a surface anisotropy constant is negative and sufficiently large in absolute value, the orientation of the unit magnetization vector perpendicular to the film plane is energetically favorable, in spite of a significant increase in the magnetostatic energy of the film. Experimentally, a rotation of the unit magnetization vector as a function of the film thickness was observed in a number of experiments with thin ferromagnetic films of iron, cobalt, and other ferromagnets [9-12].

The existence of surface magnetic anisotropy may be related to various physical origins, such as the specifics of the spin-orbit interaction on the ferromagnetic surface [13], the difference in atomic periods of various layers of a multilayer sample, or a film substrate [14], the presence of non-uniform mechanical stresses near the interface [15], etc. However, in terms of phenomenological description of magnetic phenomena [5], it is important that the energy density of surface anisotropy is concentrated in a very narrow region near the sample surface. Therefore, it is a surface contribution to the total energy which is proportional to the surface area of the film, and not to its volume. The general variational approach [5] shows that the influence of surface magnetic anisotropy on the behavior of a ferromagnet is manifested only under special boundary condition. The latter acts on the sample surface or the interface between different materials [5,16].

Since the theoretical calculation of the surface anisotropy constant $K_s$ is rather complicated [13-15], it is desirable to have reliable methods to determine this phenomenological constant experimentally. One of such methods can be the ferromagnetic resonance (FMR) [17,18], in which the film magnetization undergoes a rapid precession near a certain equilibrium position of the unit magnetization vector. The broadband FMR [19-23] can provide valuable information on the presence or absence of the energy interactions on the surface of a ferromagnetic film, or on the interfaces of ferromagnetic multilayers. The influence of surface anisotropy on the ferromagnetic resonance in a thin ferromagnetic film was previously considered in seminal papers by Rado and Weertman [24,25]. Unfortunately, no explicit formulas for the FMR frequency were actually obtained. Whereas in the present paper, the FMR frequencies have been derived for ferromagnetic film with surface anisotropy for the cases when the external magnetic field is applied perpendicularly or parallel to the film surface. Besides, various combinations of boundary conditions on the film surfaces have been studied. These results seem helpful for correct interpretation of the experimental data obtained by means of the broadband FMR in thin film structures.

## II. Out-of-plane external magnetic field

Consider a thin ferromagnetic film of thickness $L$ parallel to the XY plane and located in the domain $0 < z < L$. The unit magnetization vector of the film $\vec{\alpha}(\vec{r},t)$ satisfies the Landau-Lifshitz-Gilbert (LLG) equation

$$\frac{\partial \vec{\alpha}}{\partial t} = -\gamma \left[\vec{\alpha} \times \vec{H}_{ef}\right] + \kappa \left[\vec{\alpha} \times \frac{\partial \vec{\alpha}}{\partial t}\right], \tag{1}$$

where $\gamma$ is the gyromagnetic ratio, $\kappa$ being the phenomenological damping constant. The total effective magnetic field of the film $\vec{H}_{ef}$ takes into account the contributions due to the exchange and dipolar interactions as well as from the magnetic anisotropy energy.

$$\vec{H}_{ef} = \frac{C}{M_s}\Delta\vec{\alpha} - \frac{\partial w_a}{M_s \partial \vec{\alpha}} + \vec{H}' + \vec{H}_0. \tag{2}$$

Here $C = 2A$ is the exchange constant, $M_s$ is the saturation magnetization, $\vec{H}_0$ is the vector of the homogeneous applied magnetic field, $\vec{H}'$ is the vector of the demagnetizing field which is created by the volume and surface magnetic charges distributed in the volume and on the surface of the film, respectively. Finally, $w(\vec{\alpha})$ is the energy density of the magnetic anisotropy in the film volume.

The presence of surface magnetic anisotropy can be described [5-8] by introducing a surface interaction with the energy density per unit area

$$w_{sa} = K_s(\vec{\alpha}\vec{n})^2. \tag{3}$$

Here $K_s$ is the phenomenological surface anisotropy constant, and $\vec{n}$ is the unit vector of the external normal to the film surface. In the presence of the surface magnetic anisotropy, Eq. (3), the boundary condition for the unit magnetization vector is given by [5]

$$C\frac{\partial \vec{\alpha}}{\partial n} = 2|K_s|(\vec{\alpha}\vec{n})(\vec{n} - (\vec{\alpha}\vec{n})\vec{\alpha}). \tag{4}$$

It works on the upper and lower surfaces of the ferromagnetic film. In the absence of surface anisotropy on one or both surfaces of the film, $K_s = 0$, the usual boundary condition, $\partial \vec{\alpha}/\partial n = 0$, acts on the corresponding surface.

Suppose that the ferromagnetic film is placed in a sufficiently strong external magnetic field perpendicular to the film surface, $\vec{H}_0 = (0,0,H_0)$. In what follows we neglect for simplicity the effect of a small volume magnetic anisotropy of the film, setting $w(\vec{\alpha}) = 0$. In equilibrium, in the absence of magnetization perturbations, the unit magnetization vector is perpendicular to the film surface, $\vec{\alpha}^{(0)} = (0,0,1)$. Indeed, in the case considered the vector of effective magnetic field has only $z$ component, $\vec{H}_{ef}^{(0)} = \vec{H}_0 + \vec{H}'^{(0)} = (0,0,H_0 - 4\pi M_s)$. Therefore, the equilibrium condition for the unperturbed unit magnetization vector

$$\left[\vec{\alpha}^{(0)} \times \vec{H}_{ef}^{(0)}\right] = 0. \tag{5}$$

as well as the boundary condition (4) are satisfied.

Let us now consider small deviations of the unit magnetization vector from the equilibrium position. The first-order correction of the perturbation theory [5] to the unit magnetization vector is given by $\vec{\alpha}^{(1)} = (\alpha_x^{(1)}, \alpha_y^{(1)}, 0)$. The components of the first-order correction satisfy the equation of motion

$$\frac{\partial \vec{\alpha}^{(1)}}{\partial t} = -\gamma [\vec{\alpha}^{(1)} \times \vec{H}_{ef}^{(0)}] - \gamma [\vec{\alpha}^{(0)} \times \vec{H}_{ef}^{(1)}] + \kappa \left[ \vec{\alpha}^{(0)} \times \frac{\partial \vec{\alpha}^{(1)}}{\partial t} \right] \qquad (6)$$

Since the magnetization in the film plane is homogeneous, it is reasonable to assume that the perturbation of the magnetization depends only on the $z$ coordinate. Then the first order correction to the effective magnetic field vector is given only by the exchange interaction, so that

$$\vec{H}_{ef}^{(1)} = \left( \frac{C}{M_s} \frac{\partial^2 \alpha_x^{(1)}}{\partial z^2}, \frac{C}{M_s} \frac{\partial^2 \alpha_y^{(1)}}{\partial z^2}, 0 \right). \qquad (7)$$

Taking into account Eqs. (6), (7) and neglecting the attenuation, $\kappa \to 0$, one obtains the following equations of motion for the components of the unit magnetization vector

$$\frac{\partial \alpha_x^{(1)}}{\partial t} = \gamma \frac{C}{M_s} \frac{\partial^2 \alpha_y^{(1)}}{\partial z^2} + (\omega_m - \omega_H) \alpha_y^{(1)}; \qquad (8a)$$

$$\frac{\partial \alpha_y^{(1)}}{\partial t} = -\gamma \frac{C}{M_s} \frac{\partial^2 \alpha_x^{(1)}}{\partial z^2} - (\omega_m - \omega_H) \alpha_x^{(1)}, \qquad (8b)$$

where we denote $\omega_H = \gamma H_0$ and $\omega_m = 4\pi\gamma M_s$. The boundary conditions to Eq. (8) follow from Eq. (4)

$$C \frac{\partial \alpha_x^{(1)}(L,t)}{\partial z} = -2|K_{1s}|\alpha_x^{(1)}(L,t); \qquad C \frac{\partial \alpha_y^{(1)}(L,t)}{\partial z} = -2|K_{1s}|\alpha_y^{(1)}(L,t); \qquad (9a)$$

$$C \frac{\partial \alpha_x^{(1)}(0,t)}{\partial z} = 2|K_{2s}|\alpha_x^{(1)}(0,t); \qquad C \frac{\partial \alpha_y^{(1)}(0,t)}{\partial z} = 2|K_{2s}|\alpha_y^{(1)}(0,t). \qquad (9b)$$

Here it is assumed that the surface anisotropy constants at the upper, $K_{1s}$, and lower, $K_{2s}$, film surfaces can be different.

If one of the surface anisotropy constants is zero, for example, $K_{2s} = 0$, the solution of Eq. (8), (9) has the form of the unit magnetization vector precession

$$\alpha_x^{(1)}(z,t) = A\cos(kz)\cos(\omega t); \qquad \alpha_y^{(1)}(z,t) = B\cos(kz)\sin(\omega t). \qquad (10)$$

From the equations of motion (8) one obtains for the precession frequency

$$\omega = \omega_H - \omega_m + \gamma \frac{C}{M_s} k^2, \qquad (11)$$

whereas the boundary conditions (9) lead to the dispersion equation

$$k \tan(kL) = \frac{2|K_{1s}|}{C}. \qquad (12)$$

In the limit $kL \ll 1$ for the lowest quasi-homogeneous precession mode one obtains from Eq. (12)

$$k = \sqrt{\frac{2|K_{1s}|}{CL}}; \qquad \omega = \omega_H - \omega_m + \gamma \frac{2|K_{1s}|}{M_s L}. \qquad (13)$$

For inhomogeneous FMR modes in the limit $|K_{1s}|L/C \ll 1$ one has

$$k_n \approx \frac{\pi n}{L} + \frac{2|K_{1s}|}{C\pi n}; \qquad n = 1, 2, \ldots$$

However, if the surface anisotropy is present on both surfaces of the ferromagnetic film, the dispersion relation (12) takes the form

$$\left( k - \frac{4|K_{1s} K_{2s}|}{kC^2} \right) \tan(kL) = \frac{2(|K_{1s}| + |K_{2s}|)}{C}. \qquad (14)$$

As a result, in the limit $kL \ll 1$ the resonance frequency of the quasi-homogeneous mode equals

$$\omega = \omega_H - \omega_m + \gamma \left( \frac{2(|K_{1s}| + |K_{2s}|)}{M_s L} + \frac{4|K_{1s} K_{2s}|}{M_s C} \right). \qquad (15)$$

Eqs. (11-15) for FMR frequencies are valid in a sufficiently strong perpendicular magnetic field, under the condition that the unperturbed film magnetization is perpendicular to the film surface. They can be used to determine experimentally the values of the film surface anisotropy constants.

### III. In-plane external magnetic field

Suppose now that a sufficiently strong external magnetic field is applied parallel to the film surface, for definiteness, along the Y axis, so that $\vec{H}_0 = (0, H_0, 0)$. Then in equilibrium, in the absence of magnetization perturbation, the unit magnetization vector is parallel to the external magnetic field, $\vec{\alpha}^{(0)} = (0, 1, 0)$. Indeed, in the case under consideration the demagnetizing field is absent, $\vec{H}'^{(0)} = 0$, and the effective magnetic field vector is given by $\vec{H}_{ef}^{(0)} = \vec{H}_0 + \vec{H}'^{(0)} = (0, H_0, 0)$. Therefore, both the equilibrium equation (5) and the boundary condition (4) are satisfied.

Next, the first-order correction of the unit magnetization vector has the form $\vec{\alpha}^{(1)} = (\alpha_x^{(1)}, 0, \alpha_z^{(1)})$. The first order correction to the effective magnetic field vector contains the contributions of exchange and magneto-dipole interactions associated with the deviation of the unit magnetization vector perpendicular to the film surface

$$\vec{H}_{ef}^{(1)} = \left( \frac{C}{M_s} \frac{\partial^2 \alpha_x^{(1)}}{\partial z^2}, 0, \frac{C}{M_s} \frac{\partial^2 \alpha_z^{(1)}}{\partial z^2} - 4\pi M_s \alpha_z^{(1)} \right). \qquad (16)$$

As a result, neglecting attenuation, the equation of motion (6) takes the form

$$\frac{\partial \alpha_x^{(1)}}{\partial t} = -\gamma \frac{C}{M_s} \frac{\partial^2 \alpha_z^{(1)}}{\partial z^2} + (\omega_m + \omega_H) \alpha_z^{(1)}; \qquad (17a)$$

$$\frac{\partial \alpha_z^{(1)}}{\partial t} = \gamma \frac{C}{M_s} \frac{\partial^2 \alpha_x^{(1)}}{\partial z^2} - \omega_H \alpha_x^{(1)}. \qquad (17b)$$

The boundary conditions for the components of the unit magnetization vector are given by

$$\frac{\partial \alpha_x^{(1)}(L, t)}{\partial z} = 0; \qquad C \frac{\partial \alpha_z^{(1)}(L, t)}{\partial z} = 2|K_{1s}| \alpha_z^{(1)}(L, t); \qquad (18a)$$

$$\frac{\partial \alpha_x^{(1)}(0,t)}{\partial z} = 0; \qquad \frac{\partial \alpha_z^{(1)}(0,t)}{\partial z} = -2|K_{2s}|\alpha_z^{(1)}(0,t). \qquad (18b)$$

In the absence of the surface magnetic anisotropy, $K_{1s} = K_{2s} = 0$, the solution of Eqs. (17), (18) is similar to Eq. (10). For the FMR frequency one obtains the Kittel's relation [18]

$$\omega_n = \sqrt{\left(\omega_H + \omega_m + \gamma \frac{C}{M_s} k_n^2\right)\left(\omega_H + \gamma \frac{C}{M_s} k_n^2\right)}; \qquad k_n = \pi n/L; \qquad n = 0,1 \ldots$$

For the homogeneous precession mode, $n = 0$, it reduces to well known result [17,18] $\omega = \sqrt{\omega_H(\omega_H + \omega_m)}$.

If the surface magnetic anisotropy is present only on the upper surface of the ferromagnetic film, $K_{1s} < 0$, $K_{2s} = 0$, the solution of Eqs. (17), (18) is of the form

$$\alpha_x^{(1)}(z,t) = (A_1 \cosh(k_1 z) + A_2 \cosh(k_2 z))\cos(\omega t);$$
$$\alpha_z^{(1)}(z,t) = (B_1 \cosh(k_1 z) + B_2 \cosh(k_2 z))\sin(\omega t). \qquad (19)$$

Here the wave vectors $k_{1,2}$ and the coefficients $B_{1,2}$ are determined by the equations

$$k_{1,2}^2 = \frac{M_s}{\gamma C}\left(\omega_H + \frac{\omega_m}{2} \pm \sqrt{\omega^2 + \frac{\omega_m^2}{4}}\right); \qquad B_{1,2} = \frac{\omega_H - \frac{\gamma C}{M_s} k_{1,2}^2}{\omega} A_{1,2}.$$

In the given case, $K_{2s} = 0$, the boundary conditions (18b) for the solutions (19) are automatically satisfied, whereas the boundary conditions (18a) lead to the dispersion equation for the FMR frequency $\omega$

$$\frac{\gamma C}{M_s} k_1^2 + \frac{2|K_{1s}|}{C}\left(\omega_H - \frac{\gamma C}{M_s} k_1^2\right)\frac{\coth(k_1 L)}{k_1} = \frac{\gamma C}{M_s} k_2^2 + \frac{2|K_{1s}|}{C}\left(\omega_H - \frac{\gamma C}{M_s} k_2^2\right)\frac{\coth(k_2 L)}{k_2}. \qquad (20)$$

In the limit $k_1 L, k_2 L \ll 1$, using the series expansion $\coth(x) = 1/x + x/3 + \ldots$ for the lowest precession mode one obtains from Eq. (20) the relation

$$\omega = \sqrt{\omega_H\left(\omega_H + \omega_m - \gamma \frac{2|K_{1s}|}{M_s L}\bigg/\left(1 - \frac{2}{3}\frac{|K_{1s}|L}{C}\right)\right)}. \qquad (21)$$

Eq. (21) is valid for sufficiently small film thicknesses $L$, when dimensionless parameter $|K_{1s}|L/C < 1$. With increasing film thickness, the value of the FMR frequency $\omega$ should be found numerically from Eq. (20).

Eq. (21) for the lowest FMR frequency is used in the literature [19-21,23] by analogy with Kittel's formulas [17], but without any justification. In fact, Eq. (21) is valid only in the limit $k_1 L, k_2 L \ll 1$, and for the case when the surface anisotropy is present only on one of the film surfaces.

If the surface anisotropy is present both on the upper and lower surfaces of the ferromagnetic film, $K_{1s}, K_{2s} < 0$, the solution of Eqs. (17) has a more general form

$$\alpha_x^{(1)}(z,t) = (A_1 \cosh(k_1 z) + A_2 \sinh(k_1 z) + A_3 \cosh(k_2 z) + A_4 \sinh(k_2 z))\cos(\omega t);$$
$$\alpha_z^{(1)}(z,t) = (B_1 \cosh(k_1 z) + B_2 \sinh(k_1 z) + B_3 \cosh(k_2 z) + B_4 \sinh(k_2 z))\sin(\omega t), \qquad (22)$$

where the coefficients $B_i$ are expressed in terms of $A_i$ as follows

$$B_{1,2} = \frac{\omega_H - \frac{\gamma C}{M_s}k_1^2}{\omega} A_{1,2}; \qquad B_{3,4} = \frac{\omega_H - \frac{\gamma C}{M_s}k_2^2}{\omega} A_{3,4}.$$

Using the solutions (22), it is possible to satisfy the general boundary conditions (18). Then, from the corresponding dispersion equation in the limit $k_1 L$, $k_2 L \ll 1$ one can obtain an approximate expression for the FMR frequency

$$\omega = \sqrt{\omega_H \left(\omega_H + \omega_m - \gamma \left(\frac{2(|K_{1s}|+|K_{2s}|)}{M_s L} - \frac{4|K_{1s} K_{2s}|}{M_s C}\right)\right)}. \qquad (23)$$

As an example, Fig. 1 shows the FMR frequencies, $f = \omega/2\pi$, in the thin-film structure [2-4] Ta/CoFeB/MgO in the case when the external magnetic field $H_0$ is applied parallel to the film plane, and under the assumption that the surface anisotropy constant is nonzero only at the CoFeB/MgO interface. The solid curves in Fig. 1 are drawn according to Eq. (21).

For sufficiently small CoFeB film thicknesses this equation approximates the numerical solution of Eq. (20) with reasonable accuracy. The values of the saturation magnetization $M_s = 1250$ emu/cm$^3$, and the exchange constant $A = C/2 = 1.5 \times 10^{-6}$ erg/cm, used in the calculation were determined experimentally [22] for ferromagnetic film of the composition Co$_{40}$Fe$_{40}$B$_{20}$.

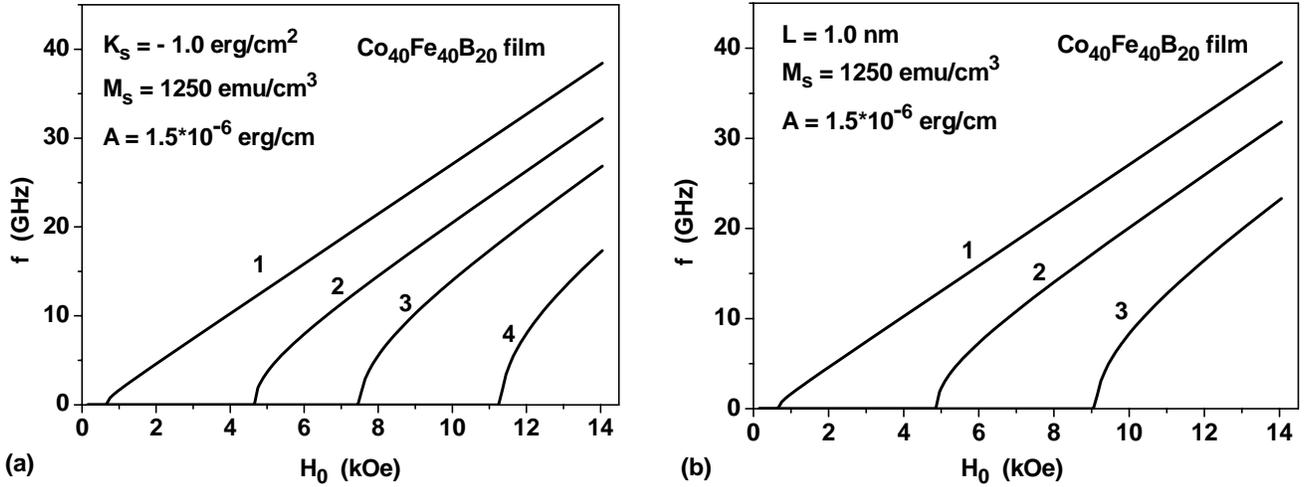

Fig. 1. a) The frequency of ferromagnetic resonance in a ferromagnetic film Co$_{40}$Fe$_{40}$B$_{20}$ as a function of in-plane magnetic field for different film thicknesses: 1) $L = 1.0$ nm; 2) $L = 0.8$ nm; 3) $L = 0.7$ nm; 4) $L = 0.6$ nm. b) The same for different values of the surface anisotropy constant at fixed thickness $L = 1.0$ nm: 1) $K_{1s} = -1.0$ erg/cm$^2$, 2) $K_{1s} = -1.25$ erg/cm$^2$, 3) $K_{1s} = -1.5$ erg/cm$^2$.

It was experimentally shown [3] that for thin-film structure Ta/CoFeB/MgO the out-of-plane magnetization exists in the film thickness range 0.6 - 1.2 nm. It can be shown also [26] that the magnetization vector becomes parallel to the film plane above the saturation field $H_s = 2|K_{1s}|/M_s L - 4\pi M_s$. This value closely coincides with the magnetic field at which the FMR frequencies, Eq. (21), become nonzero.

As Fig. 1 shows, in the presence of surface anisotropy the FMR frequencies essentially depend both on the ferromagnetic film thickness and on the value of the surface anisotropy constant. Note that Eqs. (21), (23) are valid also at film thicknesses greater than the critical value, when in the absence of applied magnetic field there is a canted or inhomogeneous micromagnetic state in the film [11,16]. The only condition is that there is a sufficiently strong in-plane magnetic field in which the unperturbed film magnetization is parallel to the film surface.

## IV. Conclusion

It this paper, it is shown that the peculiarities of the FMR spectra observed in the ultra-thin CoFeB/MgO bilayers [2-4] and similar ferromagnetic structures may be related with the presence of the surface magnetic anisotropy of appreciable value at various surfaces of the film. A proper theoretical method to describe the influence of the surface magnetic anisotropy on the film behavior is the use of the natural boundary condition that can be obtained on the basis of the general variational approach [5]. Using this technique, the FMR frequencies are obtained in this paper for a thin ferromagnetic film with surface anisotropy for the cases when the external magnetic field is applied perpendicularly or parallel to the film plane, and for various combinations of boundary conditions on the film surface. These results can be used for determination of the surface anisotropy constant of the film by means of the broadband FMR experiment.

## Acknowledgement

This work was supported by the Ministry of Education and Science of the Russian Federation in the framework of Increase Competitiveness Program of NUST «MISiS», contract № K2-2017-008.

## References

[1] S. Mangin, D. Ravelosona, J. A. Katine, M. J. Carey, B. D. Terris, E. E. Fullerton, Nature Mater. 5 (2006) 210.
[2] S. Ikeda, K. Miura, H. Yamamoto, K. Mizunuma, H. D. Gan, M. Endo, S. Kanai, J. Hayakawa, F. Matsukura, H. Ohno, Nature Mater. 9 (2010) 721.
[3] W. X. Wang, Y. Yang, H. Naganuma, Y. Ando, R. C. Yu, X. F. Han. Appl. Phys. Lett. 99 (2011) 012502.
[4] Q. L. Ma, S. Iihama, T. Kubota, X. M. Zhang, S. Mizukami, Y. Ando, T. Miyazaki, Appl. Phys. Lett. 101 (2012) 122414.
[5] W. F. Brown, Jr., Micromagnetics, Wiley-Interscience, New York - London, 1963.
[6] M. T. Johnson, P. J. H. Bloemen, F. J. A. den Broeder, J. J. de Vries, Rep. Prog. Phys. 59 (1996) 1409.
[7] H. N. Bertram, D. I. Paul, J. Appl. Phys. 82 (1997) 2439.
[8] C. A. F. Vaz, J. A. C. Bland, G. Lauhoff, Rep. Prog. Phys. 71 (2008) 056501.
[9] R. Allenspach, M. Stampanoni, A. Bischof, Phys. Rev. Lett. 65 (1990) 3344.
[10] H. Fritzsche, J. Kohlepp, H. J. Elmers, U. Gradmann, Phys. Rev. B 49 (1994) 15665.
[11] M. Speckmann, H. P. Oepen, H. Ibach, Phys. Rev. Lett. 75 (1995) 2035.
[12] M. Hehn, S. Padovani, K. Ounadjela, J. P. Bucher, Phys. Rev. B 54 (1996) 3428.
[13] H. X. Yang, M. Chshiev, B. Dieny, J. H. Lee, A. Manchon, K. H. Shin, Phys. Rev. B 84 (2011) 054401.
[14] K. H. He, S. J. Chen, J. Appl. Phys. 111 (2012) 07C109.
[15] J. I. Hong, S. Sankar, A. E. Berkowitz, W. F. Egelhoff Jr., J. Magn. Magn. Mater. 285 (2005) 359.
[16] N. A. Usov, O. N. Serebryakova, J. Appl. Phys. 121 (2017) 133905.
[17] C. Kittel, Phys. Rev. 73 (1948) 155.
[18] C. Kittel, Phys. Rev. 110 (1958) 1295.
[19] J.-M. L. Beaujour, W. Chen, A. D. Kent, J. Z. Sun, J. Appl. Phys. 99 (2006) 08N503.
[20] J-M. L. Beaujour, J. H. Lee, A. D. Kent, K. Krycka, C-C. Kao, Phys. Rev. B 74 (2006) 214405.
[21] X. Liu, W. Zhang, M. J. Carter, G. Xiao, J. Appl. Phys. 110 (2011) 033910.
[22] A. Conca, J. Greser, T. Sebastian, S. Klingler, B. Obry, B. Leven, B. Hillebrands, J. Appl. Phys. 113 (2013) 213909.
[23] A. Conca, A. Niesen, G. Reiss, B. Hillebrands, J. Phys. D.: Appl. Phys. (2018), DOI: 10.1088/1361-6463/aab5d8.
[24] G. T. Rado, J. R. Weertman, J. Phys. Chem. Solids 11 (1959) 315.
[25] G. T. Rado, Phys. Rev. B 26 (1992) 295.
[26] N. A. Usov, O. N. Serebryakova, J. Magn. Magn. Mater. 142 (2018) 453.